\documentstyle[11pt,newpasp,twoside,epsf]{article}
\def\edcomment#1{\iffalse\marginpar{\raggedright\sl#1\/}\else\relax\fi}
\reversemarginpar

\begin{document}
\title{Relativistic Binary Pulsar B1913+16: Thirty Years of Observations 
and Analysis}
\author{Joel M. Weisberg}
\affil{Dept. of Physics \& Astronomy, Carleton College, Northfield, MN}
\author{Joseph H. Taylor}
\affil{Dept. of Physics, Princeton University, Princeton, NJ}

\begin{abstract}

We describe results derived from thirty years of observations of PSR
B1913+16.  Together with the Keplerian orbital parameters,
measurements of the relativistic periastron advance and a combination
of gravitational redshift and time dilation yield the stellar masses
with high accuracy.  The measured rate of change of orbital period
agrees with that expected from the emission of gravitational
radiation, according to general relativity, to within about 0.2
percent.  Systematic effects depending on the pulsar distance and on
poorly known galactic constants now dominate the error budget, so
tighter bounds will be difficult to obtain.  Geodetic precession of
the pulsar spin axis leads to secular changes in pulse shape as the
pulsar-observer geometry changes.  This effect makes it possible to
model the two-dimensional structure of the beam.  We find that the
beam is elongated in the latitude direction and appears to be
pinched in longitude near its center.
\end{abstract}

\section{Introduction}
Pulsar B1913+16 was the first binary pulsar to be discovered (Hulse \&
Taylor 1975).  Thirty years of subsequent observations have enabled us
to measure numerous relativistic phenomena.  We have used these
measurements for fundamental tests of gravitational physics and to
place tight constraints on physical parameters of the system.  In this
paper, we provide the latest results of our observations and analysis.

\section{Observations}
The observable pulsar is a weak radio source with a flux density of
about 1~mJy at 1400~MHz.  Its observations are nearly always
sensitivity limited.  Over the years we and our colleagues have built
a number of sensitive receiver ``back ends'' for use at Arecibo,
including a swept local-oscillator system programmed to follow the
dispersed pulse, as well as filter banks and signal averagers (see
Weisberg \& Taylor 2003 for a summary).  Our most recent data have
been gathered with the Wideband Arecibo Pulsar Processors (``WAPPs''),
which for PSR B1913+16 achieve $13~\mu$s time-of-arrival measurements
in each of four 100 MHz bands, using 5-minute integrations.  In
addition to the timing observations, we have used the Princeton Mark
IV coherent de-dispersing system (Stairs et al. 2000) to measure
average pulse shapes and polarization over the last five years.  A
major improvement in data acquisition for this experiment was made in
1981, and data taken since then have much higher quality than the
earlier observations.  The results reported here are based largely on
data gathered from 1981 through 2003.

\section{Relativistic Timing Analysis}

Non-relativistic analysis of arrival time data from this system can
yield five orbital parameters: the projected semimajor axis of the
pulsar orbit $a_p \sin i$, orbital eccentricity $e$, epoch of
periastron $T_0$, orbital period $P_b$, and argument of periastron
$\omega_0$.  Relativistic effects lead to three additional
measurables: the mean rate of advance of periastron
$\langle\dot{\omega}\rangle$, gravitational redshift and time-dilation
parameter $\gamma$, and orbital period derivative
$\dot{P}_b$. Measured pulse times of arrival calculated for each five
minutes of observation serve as the input data for program TEMPO
(http://pulsar.princeton.edu/tempo).  This program fits a model with
eighteen parameters (eight orbital quantities plus ten astrometric and
spin parameters) to the data, using the timing model of Damour \&
Deruelle (1985, 1986).  Fitted values for the orbital parameters are
listed in Table 1, with uncertainties in the last shown in
parentheses.

\begin{table}
\begin{center}
\caption{Measured Orbital  Parameters for B1913+16 System}
\medskip
\begin{tabular}{ll}
\tableline
Fitted Parameter		        &Value	    \\
\tableline
$a_p \sin$ i (s)	\dotfill	& 2.3417725 (8)    \\
e 		\dotfill		& 0.6171338 (4)   \\
T$_0$  (MJD)	\dotfill    	 & 52144.90097844 (5)  \\
P$_b$  (d)	\dotfill           & 0.322997448930 (4) \\
$\omega_0$ (deg)\dotfill		  & 292.54487 (8)       \\
$\langle\dot{\omega}\rangle$ (deg/yr)\dotfill	& 4.226595 (5)      \\
$\gamma$ (s)\dotfill				& 0.0042919 (8)      \\
$\dot{P}_b$  ($10^{-12}$ s/s)\ldots	       & $-$2.4184 (9)      \\
\tableline
\tableline
\end{tabular}
\end{center}
\end{table}

The pulsar orbit is fully specified (up to an unknown rotation about
the line of sight) by the first seven parameters listed in Table~1.
Other orbital quantities such as inclination, masses of the stellar
components, and the semimajor axes, may be derived from these seven;
Taylor \& Weisberg (1982) provide the relevant formulas.  For example,
the masses of the pulsar and companion are $m_p=1.4414\pm0.0002$ and
$m_c=1.3867\pm0.0002$ solar masses, respectively.  (Note that in order
to express the masses in grams, a value would need to be introduced
for the Newtonian gravitational constant $G$.  The uncertainty in $G$
is comparable to our quoted uncertainties in $m_p$ and $m_c$.)  As
described below, the eighth measured orbital parameter, $\dot{P}_b$,
overdetermines the system dynamically and thus provides a test of
gravitation theory.

\subsection{Emission of Gravitational Radiation}

According to general relativity, a binary star system should emit
energy in the form of gravitational waves.  The loss of orbital energy
results in shrinkage of the orbit, which is most easily observed
as a decrease in orbital period.  Peters \& Matthews (1963) showed
that in general relativity the rate of period decrease is given by
\begin{eqnarray}
\dot{P}_{b,GR}=-\frac{192\ \pi \ G^{5/3} } {5 \ c^5}  \
\left(\frac{P_b}{2\pi}\right)^{-5/3} \  (1-e^2)^{-7/2} \   \times \\
\nonumber \left(1 + \frac{73}{24} e^2 + \frac{37}{96} e^4\right)
 \ m_p \ m_c \ (m_p + m_c)^{-1/3}.
\end{eqnarray}

Note that except for Newton's constant $G$ and the speed of light $c$,
all quantities on the right hand side of Eq.~(1) have measured values
listed in Table 1, or, in the case of the component masses, are
derivable from those quantities.  The predicted orbital period
derivative due to gravitational radiation computed from
Eq.~(1) is $\dot{P}_{b,GR}=-(2.40242\pm0.00002)\times 10^{-12}$ s/s.

Comparison of the measured $\dot{P}_{b}$ with the theoretical value
requires a small correction, $\dot{P}_{b,Gal}$, for relative
acceleration between the
solar system and binary pulsar system, projected onto the line of
sight (Damour \& Taylor 1991).  This correction
is applied to the measured $\dot{P}_{b}$
to form a ``corrected'' value $\dot{P}_{b,corrected} = \dot{P}_{b} -
\dot{P}_{b,Gal}$. The correction term depends on several rather
poorly known quantities, including the distance and proper
motion of the pulsar and the radius of the Sun's galactic orbit.  The
best currently available values yield
$\dot{P}_{b,Gal}=-(0.0128\pm0.0050)\times10^{-12}$ s/s, so that
$\dot{P}_{b,corrected}=(2.4056\pm0.0051)\times 10^{-12}$ s/s.  Hence
\begin{equation}
\frac{{\dot{P}_{b,corrected}}}{\dot{P}_{b,GR}}=1.0013\pm0.0021,
\end{equation}
and we conclude that the measured orbital decay is consistent at the
$(0.13\pm0.21) \%$ level with the general relativistic prediction for
the emission of gravitational radiation.  The observed and theoretical
orbital decays are compared graphically in Figure 1.

\begin{figure}
\plotfiddle{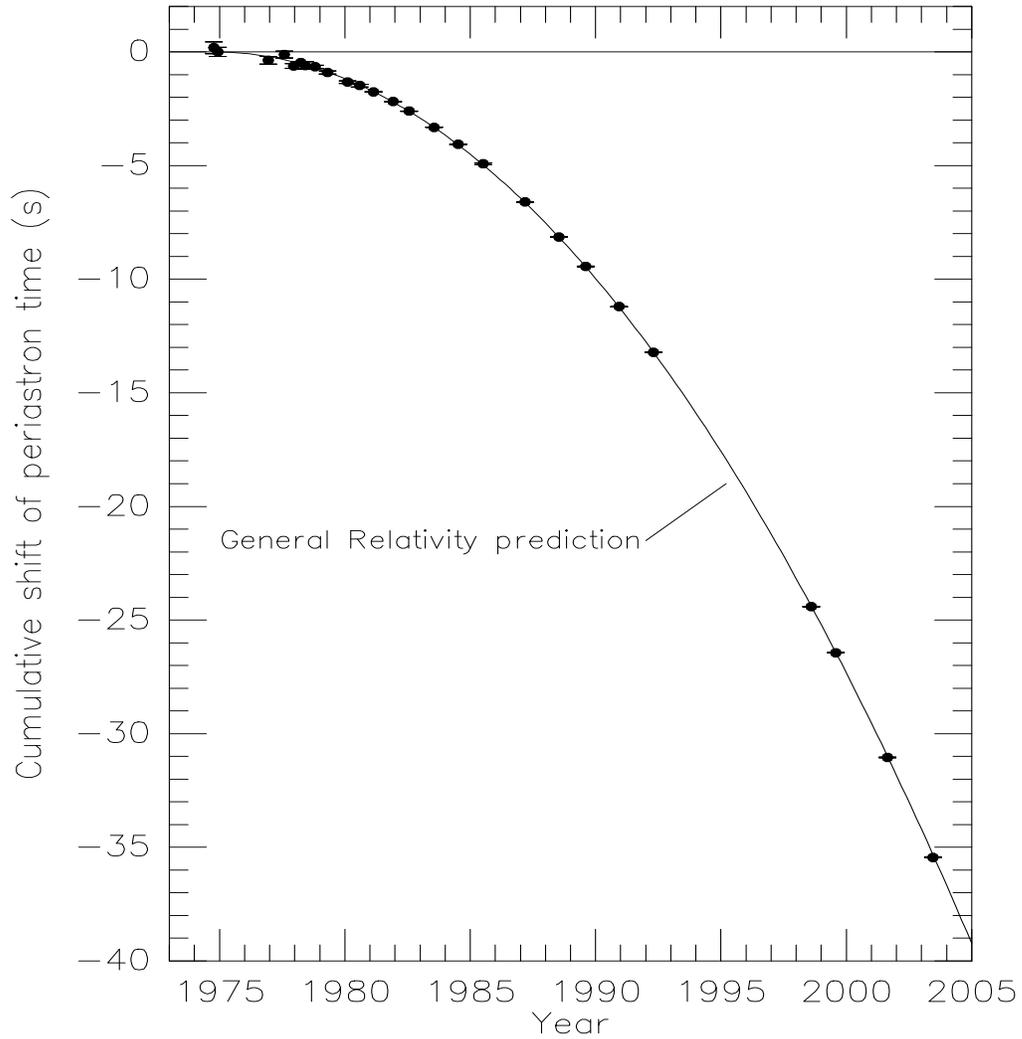}{3.5 in}{0}{70}{60}{-220}{-40}
\caption{Orbital decay of PSR B1913+16. The data points indicate the
observed change in the epoch of periastron with date while the
parabola illustrates the theoretically expected change in epoch for a
system emitting gravitational radiation, according to general
relativity.}
\end{figure}

Accuracy of the test for gravitational radiation damping is now
dominated by the uncertainty in the galactic acceleration term.  Work
now underway should lead to improved accuracy of the pulsar proper
motion, and the Sun's galactocentric distance may be better known in
the future.  However, we see little prospect for a significant
improvement in knowledge of the pulsar distance.  Consequently, it
seems unlikely that this test of relativistic gravity will be improved
significantly in the foreseeable future.

\section{Geodetic Precession: Mapping the Emission Beam}

Relativistic spin-orbit coupling causes the pulsar's spin axis to
precess (Damour \& Ruffini 1974; Barker \&O'Connell 1975a,b).  In the
PSR B1913+16 system this so-called ``geodetic'' precession has a
period of about 300~y.  The resulting change of aspect with respect to
the line of sight should cause a secular change in pulse shape.
Weisberg, Romani, \& Taylor (1989) reported shape changes at 1400~MHz
and attributed them to the line of sight moving across the middle of a
hollow-cone beam.  Kramer (1998) found that the separation between the
two principal pulse components began to shrink in the mid-1990s,
suggesting that the line of sight had continued to drift across the
conal beam, moving farther from its center.  Kramer fitted a model to
these data, which indicated that the pulsar spin and orbital angular
momenta are misaligned by $\sim20\deg$ and that the beam will no
longer intersect our line of sight after the year 2025. Weisberg \&
Taylor (2002) confirmed these results and found that the conal beam
appears to be elongated in the direction parallel to the spin axis;
they also noted that the beam appears to be ``hourglass-shaped,'' or
pinched in longitude near its center (see Fig.~2). Kramer (2002)
argued that these observations could also result from the line of
sight precessing away from the center of a circular emission cone and
onto a core pencil beam that is offset below the center and toward the
the trailing part of the cone.

\begin{figure}
\plotfiddle{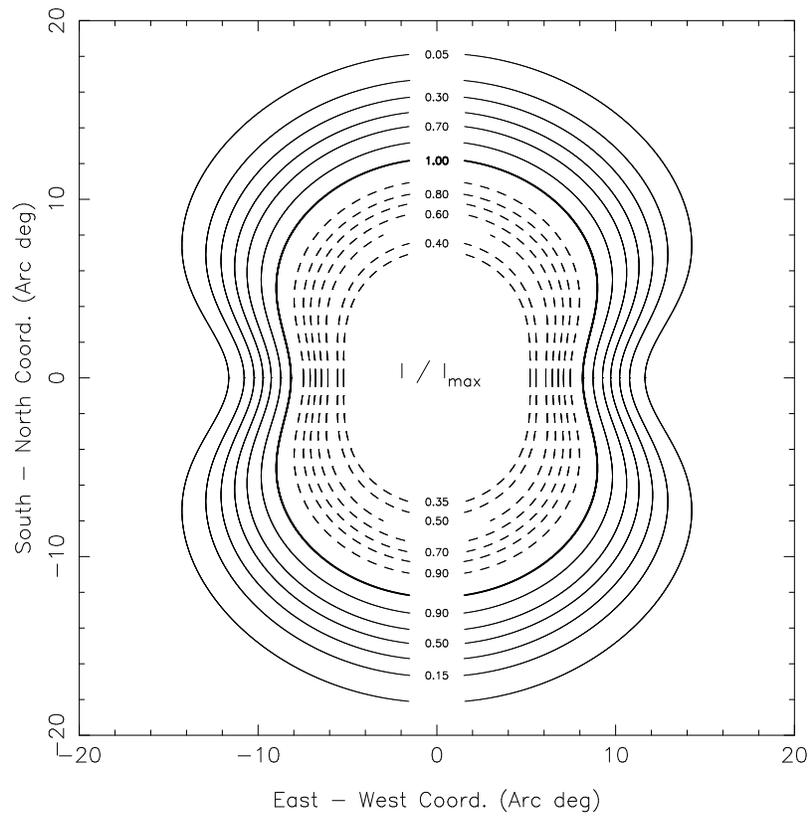}{4.5 in}{270}{60}{60}{-220}{360}
\caption{Hourglass-shaped conal beam model. The model was fitted to the even
components of 1400 MHz profiles from 1981 to 2003.}
\end{figure}

We divided all of our 1400 MHz profiles into components that are
respectively even and odd about the profile center.  We then fitted
the even components to a model that allows for a noncircular conal
beam (Weisberg \& Taylor 2002).  With more data now available, we find
that the model parameters have changed little from earlier solutions.
The current model is illustrated in Figure~2 in the form of
equal-intensity contours of the beam.

Rankin (1983) showed that in most pulsars core emission becomes more
prominent at low frequencies.  We have observed PSR B1913+16 at the
lower frequency of 430 MHz at several epochs, and three of the
resulting pulse profiles are shown in Figure 3.  A core component is
quite prominent in the data taken in 1980-81, but it faded very
significantly between 1980 and 1998 and was nearly gone by 2003.  This
behavior further supports our model in which the line of sight is
precessing away from the axis of a centered pencil beam plus a 
longitude-pinched cone
elongated in the latitude direction.

Even with 30 years of observations, only a small portion of the
north-south extent of the emission beam has been observed.  As a
consequence, our model is neither unique nor particularly robust.  The
north-south symmetry of the model is assumed, not observed, since the
line of sight has fallen on the same side of the beam axis throughout
these observations.  Nevertheless, accumulating data continue to
support the principal features noted above.

\begin{figure}
\plotone{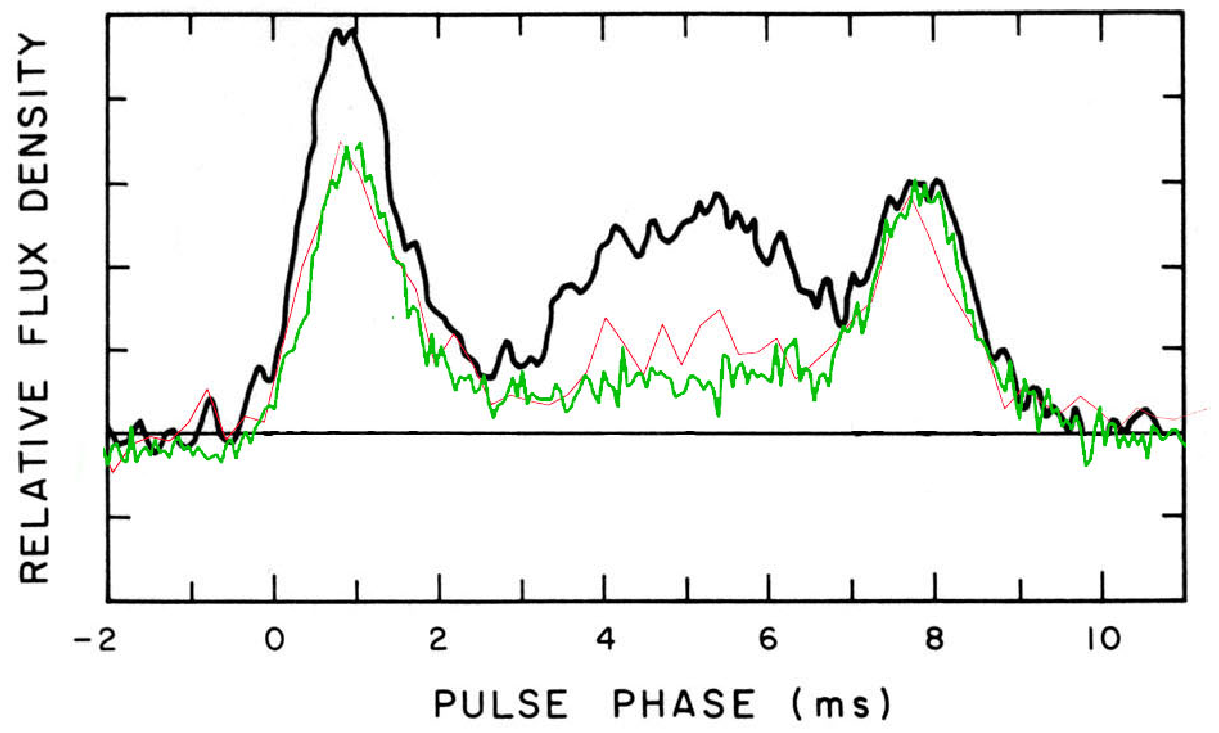}
\caption{Profiles of PSR B1913+16 at 430 MHz at three epochs.  The core
component declines from 1980 to 1998 through 2003, indicating that the line
of sight is precessing away from the axis of a {\it{centered}} core pencil
beam.}
\end{figure}

\end{document}